\documentclass[a4paper,12pt, epsfig]{article}
\usepackage{epsfig}
\usepackage{amssymb}
\usepackage{amsfonts}
\usepackage{graphicx,psfrag}

\newskip\humongous \humongous=0pt plus 1000pt minus 1000pt

\newif\ifdtup

\catcode`\@=11

\@addtoreset{equation}{section}
\def\theequation{\thesection.\arabic{equation}}

\def\@normalsize{\@setsize\normalsize{15pt}\xiipt\@xiipt
\abovedisplayskip 14pt plus3pt minus3pt%
\belowdisplayskip \abovedisplayskip
\abovedisplayshortskip \z@ plus3pt%
\belowdisplayshortskip 7pt plus3.5pt minus0pt}

\def\small{\@setsize\small{13.6pt}\xipt\@xipt
\abovedisplayskip 13pt plus3pt minus3pt%
\belowdisplayskip \abovedisplayskip
\abovedisplayshortskip \z@ plus3pt%
\belowdisplayshortskip 7pt plus3.5pt minus0pt
\def\@listi{\parsep 4.5pt plus 2pt minus 1pt
     \itemsep \parsep
     \topsep 9pt plus 3pt minus 3pt}}

\relax

\catcode`@=12

\catcode`\@=11

\def\section{\@startsection{section}{1}{\z@}{3.5ex plus 1ex minus
   .2ex}{2.3ex plus .2ex}{\large\bf}}

\def\thesection{\arabic{section}}
\def\thesubsection{\arabic{section}.\arabic{subsection}}

\def\appendix{\setcounter{section}{0}
 \def\thesection{Appendix \Alph{section}}
 \def\thesubsection{\Alph{section}.\arabic{subsection}}
 \def\theequation{\Alph{section}.\arabic{equation}}}

\def\SymBoxes#1#2#3#4{\newdimen\un@t \un@t#3%
\raisebox{#1}{\rule{#2\un@t}{#4}\hskip-#2\un@t
\@tempdimb\un@t \advance\@tempdimb by-#4\@tempcntb#2\relax%
\@whilenum{\@tempcntb>0}\do{
\rule{#4}{\un@t}\hskip\@tempdimb \advance\@tempcntb by\m@ne}%
\hskip-#2\un@t \rule[\un@t]{#2\un@t}{#4}%
\rule[\un@t]{#4}{#4}\hskip-#4
\rule{#4}{\un@t}}\hskip-#4}                

\begin{document}

\newcommand{\beq}{\begin{equation}}
\newcommand{\eeq}{\end{equation}}
\newcommand{\bea}{\begin{eqnarray}}
\newcommand{\eea}{\end{eqnarray}}
\newcommand{\beas}{\begin{eqnarray*}}
\newcommand{\eeas}{\end{eqnarray*}}
\newcommand{\defi}{\stackrel{\rm def}{=}}
\newcommand{\non}{\nonumber}
\newcommand{\bquo}{\begin{quote}}
\newcommand{\enqu}{\end{quote}}
\renewcommand{\(}{\begin{equation}}
\renewcommand{\)}{\end{equation}}
\def\de{\partial}
\def\Tr{ \hbox{\rm Tr}}
\def\H{ \hbox{\rm H}}
\def\HE{ \hbox{$\rm H^{\it H}_{even}$}}
\def\HO{ \hbox{$\rm H^{\it H}_{odd}$}}
\def\K{ \hbox{\rm K}}
\def\Im{ \hbox{\rm Im}}
\def\Ker{ \hbox{\rm Ker}}
\def\const{\hbox {\rm const.}}
\def\o{\over}
\def\im{\hbox{\rm Im}}
\def\re{\hbox{\rm Re}}
\def\bra{\langle}\def\ket{\rangle}
\def\Arg{\hbox {\rm Arg}}
\def\Re{\hbox {\rm Re}}
\def\Im{\hbox {\rm Im}}
\def\exo{\hbox {\rm exp}}
\def\diag{\hbox{\rm diag}}
\def\longvert{{\rule[-2mm]{0.1mm}{7mm}}\,}
\def\a{\alpha}
\def\dag{{}^{\dagger}}
\def\tq{{\widetilde q}}
\def\p{{}^{\prime}}
\def\W{W}
\def\N{{\cal N}}
\def\hsp{,\hspace{.7cm}}
\newcommand{\C}{\ensuremath{\mathbb C}}
\newcommand{\Z}{\ensuremath{\mathbb Z}}
\newcommand{\R}{\ensuremath{\mathbb R}}
\newcommand{\rp}{\ensuremath{\mathbb {RP}}}
\newcommand{\cp}{\ensuremath{\mathbb {CP}}}
\newcommand{\vac}{\ensuremath{|0\rangle}}
\newcommand{\vact}{\ensuremath{|00\rangle}                    }
\newcommand{\oc}{\ensuremath{\overline{c}}}
\begin{titlepage}
\begin{flushright}
ULB-TH/06-28\\
KUL-TF-06/31\\
hep-th/0611218\\
\end{flushright}
\bigskip
\def\thefootnote{\fnsymbol{footnote}}

\begin{center}
{\large {\bf
Twisted Homology
 } }
\end{center}

\bigskip
\begin{center}
{\large   Andr\'es Collinucci${}^1$\footnote{\texttt{andres.collinucci@fys.kuleuven.be}} and Jarah Evslin${}^2$\footnote{\texttt{ jevslin@ulb.ac.be}}}\\
\end{center}

\renewcommand{\thefootnote}{\arabic{footnote}}

\begin{center}
${}^1$\em {Institute for Theoretical Physics, K.U. Leuven,\\ Celestijnenlaan 200D, B-3001 Leuven, Belgium}\\

\vspace{1em}
{${}^2$\em  { International Solvay Institutes,\\
Physique Th\'eorique et Math\'ematique,\\
Statistical and Plasma Physics C.P. 231,\\
Universit\'e Libre
de Bruxelles, \\ B-1050, Bruxelles, Belgium\\}}

\end{center}

\noindent
\begin{center} {\bf Abstract} \end{center}
\noindent
D-branes are classified by twisted K-theory.  Yet twisted K-theory is often hard to calculate.  We argue that, in the case of a compactification on a simply-connected six manifold, twisted K-theory is isomorphic to a much simpler object, twisted homology.  Unlike K-theory, homology can be twisted by a class of any degree and so it classifies not only D-branes but also M-branes.  Twisted homology classes correspond to cycles in a certain bundle over spacetime, and branes may decay via Kachru-Pearson-Verlinde transitions only if this cycle is trivial.   We provide a spectral sequence which calculates twisted homology, the $k$th step treats D$(p-2k)$-branes ending on D$p$-branes.

\vfill

\end{titlepage}
\bigskip

\hfill{}
\bigskip


\setcounter{footnote}{0}
\section{Introduction}

\noindent
It has been nearly a decade since Minasian and Moore \cite{MM} proposed that D-branes in type II string theories are classified by K-theory.  Since then the proposal has been generalized to a number of different theories, from two-dimensional topological theories \cite{MS} to exactly solvable conformal field theories with target spaces of arbitrary dimension to theories with orientifolds.  The original K-theory proposal fails in the presence of a topologically nontrivial NS $H$ field \cite{Kapustin} and needs to be replaced with Rosenberg's \cite{Rosenberg} twisted K-theory \cite{BM}.  Twisted K-theory has provided the most successful model to date of conserved RR charges.  However this success comes with a price, in general twisted K-theory classes are difficult to compute.  Even once they are computed, given a representative of a fixed twisted K-theory class it is difficult to extract the physical data that describes the corresponding D-brane, such as its embedding or its worldvolume gauge field.

In this note we will describe a less general classification scheme, twisted homology, which we argue is isomorphic to twisted K-theory in the case of a simply-connected six manifold  \cite{Kreview}.  We will present evidence that this isomorphism continues to hold for all orientable, $spin^c$ six-manifolds.  Thus twisted homology is sufficient for the backgrounds of interest in string phenomenology.  However, unlike twisted K-theory, we will see that given a D-brane's embedding and worldvolume gauge field one can easily determine the corresponding twisted homology class.  This is a companion paper to Ref.~\cite{Luca}, in which twisted homology is presented using a framework similar to that of the generalized submanifolds of Ref.~\cite{Gualtieri} and is applied to the theory of calibrations.  That perspective of twisted homology was introduced in Ref.~\cite{Kreview}, although the equivalence condition dates back to Refs.~\cite{Paul,Luca1,Luca2}.

Homology, can be twisted by a cohomology class of any degree.  On the other hand, a topological realization of twisted K-theory is only known when the twisting class is a 3-class, which in string theory is either the NSNS $H$ field \cite{BM} or the 3-form RR field strength in IIB, for example in the twisted K-theory classification of branes on the conifold in Refs.~\cite{mecascade,Kreview}.  This versatility means that twisted homology with a degree 4 twist may be used to classify branes in M-theory.

Given a twist $H$ of fixed degree $k$, we present a geometrical way to visualize the twisted homology of a spacetime $M$.  One may construct a bundle $Q$ over $M$ with characteristic class $H$.   One may represent any twisted homology class of $M$ as an ordinary homology class of the bundle.  This realization has several applications.  

First, branes carry a conserved charge if they wrap a nontrivial homology class in $Q$.  Thus, branes may decay via dimension-changing processes like those of Refs.~\cite{KPV} and \cite{MMS} only if they wrap a cycle which is trivial in $Q$, but whose projection onto the spacetime $M$ may be topologically nontrivial.  We also find that a D-brane wrapping a cycle $N$ in $M$ suffers from a Freed-Witten anomaly \cite{FW} if and only if the bundle $Q$ is nontrivial over $N$, so that there is no section over $N$ which defines the cycle in $Q$ wrapped by the brane.  In other words, a brane that wraps a cycle in $Q$ automatically is Freed-Witten anomaly-free.  

In fact, unlike the simple quotient of the kernel of multiplication by $H$ by its image, twisted homology also knows about some two-step anomalies, which are classical versions of a phenomenon discovered in the SU(3) WZW model and described in Ref.~\cite{MMS}.  In two-step anomalies a D$p$-brane suffers from a Freed-Witten anomaly which is cured by a D$(p-2)$-brane insertion.  Unlike the usual Freed-Witten case, the boundary of the D$(p-2)$-brane is contractible in the total spacetime $M$, although not on the worldvolume of the D$p$-brane, and so instead of being semi-infinite it closes off.  A two-step anomaly occurs when, in the process of closing off, it wraps some $H$ flux, and so it also becomes anomalous.  This second anomaly is cured by the insertion of a semi-infinite D$(p-4)$-brane.  This process, as well as arbitrary numbers of iterations, renders the original D$p$-branes inconsistent and the D$(p-4)$-branes unstable \cite{MMS}.  These inconsistent D$p$-branes represent neither twisted K-theory nor twisted homology classes, while the unstable D$(p-4)$-branes carry zero units of twisted homology and twisted K-theory charge.  On the other hand, all of these branes carry nontrivial homology charge, and also nontrivial charge under the quotient of the kernel of $H$ by its image.  Thus twisted homology and twisted K-theory capture the physics of these processes better than the other two models.  

As another application we find a formula that lets one calculate the twisted homology of the spacetime $M$.  We have argued that the twisted homology of $M$ is just a subset of the homology of $Q$.  We will see that in certain cases the twisted homology of $M$ is a particular quotient of the ordinary homology of a subbundle of $Q$ by the ordinary homology of $M$.  This is somewhat complicated by the fact that the ordinary homology of $M$ is not always a subgroup of the ordinary homology of this subbundle, instead it will be necessary to divide it into two pieces and quotient by them one at a time, as we will describe.

We begin in Sec.~\ref{defsez} with a review of simplices, chains and the cap product in algebraic topology, which we combine to define the twisted boundary operator $\partial_H$.  We define twisted homology to be the quotient of the kernel of $\partial_H$ by its image.  We describe the conditions under which $\partial_H$ is nilpotent, and the twisted homology is defined.  Then in Sec.~\ref{brane} we interpret $\partial_H$ as the operator that takes a brane to a cycle that measures its anomaly, and we argue that twisted homology classifies consistent D-branes up to dynamical processes, some of which generalize \cite{KPV}.  As an example we calculate and interpret the twisted homology of the three-sphere.  We use this interpretation of twisted homology to construct a spectral sequence that calculates twisted homology from untwisted homology.  The later steps in this spectral sequence describe multistep processes in which, for example, a brane augments its dimension multiple times before decaying.  

Finally in Sec.~\ref{sfere} we introduce an auxiliary bundle over spacetime, and we claim that the twisted homology classes in spacetime are just ordinary homology classes in the total space of the bundle.  The bundle is useful, because, unlike homology classes in spacetime, homology classes in the bundle correspond to conserved charges.  For example, a brane can only decay if it wraps a trivial cycle in the total space of the bundle.  We use this construction, along with the Gysin sequence, to prove that, under certain conditions, the twisted homology of $M$ is the quotient of the ordinary homology of the total space of the bundle by the ordinary homology of $M$.  Finally we apply this technology to calculate the twisted homology of a lens space using the spectral sequence and also using the bundle.

\section{Definitions} \label{defsez}

\subsection{Some Background Material}

We will define twisted homology using simplices.  A $k$-simplex $\sigma$ is a $k$-dimensional pyramid.  Its faces are $(k-1)$-dimensional pyramids, and so are $(k-1)$-simplices.  In algebraic topology, one often describes a space $M$ by defining a simplicial decomposition, which is a set of simplices and maps of the simplices into the spacetime such that every point in $M$ is in the image of some simplex, the intersection of any number of simplices has to be another simplex in the set and every face of a simplex in the set is another simplex in the set.  

One can construct formal sums $C$ of simplices weighted by, for example, the integers $\{n_i\}$
\beq
C=\sum_i n_i \sigma_i.
\eeq
These formal sums are called chains.  They are similar to submanifolds, but they come with weights, which we will identify with the charges of the D-branes that wrap them.  If all of the simplices with nonzero weight in a chain are of dimension $k$, then the chain is called a $k$-chain, and intuitively it is similar to a $k$-submanifold.  The set of $k$ chains forms an abelian group under addition of the coefficients, which we will denote $C_k$.  Like submanifolds, there is a natural notion of a boundary map $\partial$, which takes each $k$-simplex to the sum of its $(k-1)$-simplex faces, weighted by plus or minus one depending on a choice of orientations. This definition can be extended to general chains by imposing that it acts linearly, therefore $\partial$ takes a $k$-chain to a $(k-1)$-chain.

A $k$-chain which is in the kernel of $\partial$ is said to be a $k$-cycle, whereas a $k$-chain which is in the image of the boundary map is said to be a $k$-boundary.  The untwisted $k$th homology group is the quotient of the $k$-cycles by the $k$-boundaries
\beq
\H_k(M)=\frac{\Ker(\partial:C_k\longrightarrow C_{k-1})}{\im(\partial:C_{k+1}\longrightarrow C_{k})}.
\eeq
To extend this definition to twisted homology, we will need to define the space of allowed twists, which are cocycles.

A cochain $\gamma$ is a choice of an integer $n^i$ for each simplex $\sigma_i$.  This definition can be extended to a map on chains by demanding that the map be linear.  If the $n^i$'s are only nonzero for $k$-chains, then the cochain $\gamma$ is said to be a $k$-cochain.  We will call the space of $k$-cochains $C^k$. 

One can define a coboundary map $\delta$ on the cochains by demanding that for any chain $C$
\beq
\gamma(\partial C)=\delta\gamma(C).
\eeq
Notice that if $\gamma$ is a $k$-cochain then $\delta\gamma$ is a $(k+1)$-cochain, in fact $\delta$ is just the transpose of $\partial$.  As above, we define $k$-cocycles to be $k$-cochains in the kernel of $\delta$ and $k$-coboundaries to be $k$-cochains in the image of $\delta$.  Then integral cohomology is just the quotient of the cocycles by the coboundaries
\beq
\H^k(M)=\frac{\Ker(\delta:C^k\longrightarrow C^{k+1})}{\im(\delta:C^{k-1}\longrightarrow C^{k})}.
\eeq
Our twists will be cocycles, and so, up to coboundaries, will be classified by integral cohomology.

Finally we need to define another action of cochains on chains, which generalizes the definition of cochains as maps from the space of chains to the integers.  The cap product $\cap$ of a $j$-chain $c_j$ and a $k$-cochain $c^k$ is a $(j-k)$-chain $c_{j-k}$.  In the perhaps more familiar language of differential forms one may imagine that $c_j$ is a $j$-dimensional submanifold wrapped by a brane and $k$ is a $k$-form in the bulk which we pull back onto the brane's worldvolume.  Then $c_{j-k}$ is the Poincar\'e dual of this form in the brane's worldvolume, which is a codimension $k$ submanifold in the worldvolume.  

Formally the cap product of $c^k$ with a single $j$-simplex $\sigma_j$ is defined as follows.  Label the vertices of $\sigma_j$ by integers from $0$ to $j$. Consider the $k$-simplex $\sigma_k$ which is the face of $\sigma_j$ whose vertices are numbers $0$ to $k$.  Now we have a $k$-cochain and a $k$-simplex, recalling that a $k$-cochain is just a choice of an integer for each $k$-simplex, we obtain an integer $n$.  The cap product of $c^k$ and $\sigma_j$ is the $(j-k)$-simplex $\sigma_{j-k}$ which is the face of $\sigma_j$ with vertices numbered from $k$ to $j$, multiplied by the integer $n$
\beq
c^k\cap\sigma_j=c^k(\sigma_k)\sigma_{j-k}=n\sigma_{j-k}.
\eeq
This defines the cap product of a cochain and a simplex.  This definition can be extended to a cochain and a chain by imposing that it be linear.  

If $c^k$ is a $k$-cochain and $c_j$ is a $j$-chain then
\beq
\partial(c_j\cap c^k)=(-1)^k(\partial c_j\cap c^k-c_j\cap\delta c^k)
\eeq
and so the boundary operator is a derivation with respect to the cap product.  This implies that the cap product may be extended to homology and cohomology classes.  We will be interested instead in extending it from cochains $C^k$ to cohomology classes $\H^k$, but not in extending from chains $C_j$ to homology classes $\H_j$
\beq
\H^k\cap C_j\subset C_{j-k}.
\eeq
We will usually consider the cap product to be an action of cohomology classes on chains, intuitively the cap product with the $H$ flux takes a D$p$-brane worldvolume and gives the codimension 3 submanifold where there is a magnetic monopole with respect to the worldvolume $U(1)$ gauge field.

\subsection{Twisted Homology}
\begin{figure}[ht]
\begin{center}
\leavevmode
\epsfxsize 11   cm
\epsffile{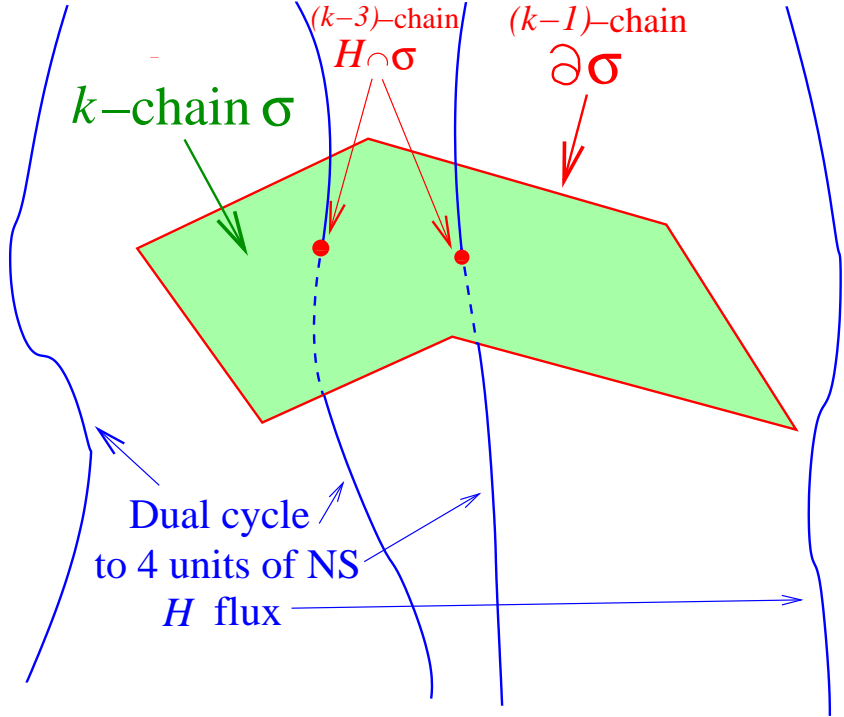}    
\end{center} 
\caption{The action of $\partial_H$ on a $k$-chain $\sigma$ is sketched.  The twist $H$ is depicted using its dual (dim$(M)$-3)-cycle, which is the cap product of $H$ with the fundamental class of the spacetime $M$. The intersection of this cycle with the chain $\sigma$ is the $(k-3)$-cycle $H\cap\sigma$.  $\partial_H\sigma$ is the sum of $H\cap\sigma$ and the boundary of $\sigma$, and is represented by everything red.}
\label{bordo} 
\end{figure}

At last we may define the twisted boundary operator
\beq
\partial_H=\partial+H\cap.
\eeq
Its action on a chain $\sigma$ is illustrated in Fig.~\ref{bordo}.  Chains in the kernel of $\partial_H$ are said to be twisted cycles and those in its image are twisted boundaries.  $\partial$ of a $k$-chain is a $(k-1)$-chain, and if $H$ is a 3-cocycle then $H\cap$ of a $k$-chain is a $(k-3)$-chain.  Therefore, $\partial_H$ of a $k$ chain is the sum of a $(k-1)$-chain and a $(k-3)$-chain.  In particular, it is possible that $H\cap$ on a $k$-chain will cancel $\partial$ on a $(k-2)$-chain, and so some twisted cycles will necessarily be of mixed dimension.  However even and odd dimensions are not mixed, the twisted boundary operator maps even chains to odd chains and odd chains to even chains.  We will denote the space of even chains and odd chains by $C_E$ and $C_O$ respectively
\beq
C_E=\oplus_i C_{2i}\hsp C_O=\oplus_i C_{2i+1}.
\eeq
We will assemble these into even an odd twisted homology classes, which will classify RR charges in IIA and IIB respectively.

Twisted homology is the quotient of the twisted cycles by the twisted boundaries
\beq \label{omrit}
\HE(M)=\frac{\Ker(\partial_H:C_E\rightarrow C_O)}{\im(\partial_H:C_O\rightarrow C_E)},\ \ \ \ 
\HO(M)=\frac{\Ker(\partial_H:C_O\rightarrow C_E)}{\im(\partial_H:C_E\rightarrow C_O)}.
\eeq
Notice that if $H=0$ then twisted even or odd homology is just the direct sum of all of the untwisted even or odd homology classes.

Twisted homology may be defined for any cocycle $H$, it need not be of degree 3.  However the twisted boundary operator only preserves the parity of the degree of the chain if $H$ is of odd degree.  If $H$ is purely of degree $k$ then $\partial_H$ preserves the degree of a chain modulo $k-1$.  For example, one may use homology twisted by the M-theory 4-form field strength $G_4$ to classify networks of M5-branes, M2-branes and M2-brane worldvolume instantons by the second twisted homology group $\H_2^{G_4}$, whose representatives are chains of degree $2$ modulo $3$.  The twist may also be of degree 2, in which case no grading is preserved and there is a single twisted homology group.

Similarly one may define twisted cohomology, using the twisted coboundary operator $\delta_H$, which is defined to be the cup product $\cup$ with $H$ plus the coboundary operator $\delta$.  The cup product is a multiplicative product on cochains, generalizing the wedge product on differential forms, that takes a $j$-cochain and a $k$-cochain to a $(j+k)$-cochain, and satisfies a compatibility condition with the cap product
\beq
A\cap(B\cap C)=(A\cup B)\cap C \label{prop}
\eeq
where $A$ and $B$ are cochains and $C$ is a chain.  Twisted cohomology has already appeared in the physics literature, as, if one uses real coefficients instead of integers, it is isomorphic to the tensor product of twisted K-theory with the real numbers via an isomorphism known as the twisted Chern character.  

Twisted cohomology and twisted homology are Poincar\'e dual, the duality is as usual realized by taking the cap product with the fundamental class $[M]$ of $M$ which exists when $M$ is orientable and compact.  As a check on this duality, we will show that twisted cocycles are dual to twisted cycles.  If $C$ is a twisted cocycle then by definition
\beq
0=\delta_H C=\delta C + H\cup C
\eeq
and so
\bea
\partial_H (C\cap [M])&=&\partial(C\cap [M])+H\cap(C\cap [M])\\
&=&(\delta C\cap [M])+(H\cup C)\cap [M]=(\delta_H C)\cap [M] =0\nonumber
\eea
where we have used Eq.~(\ref{prop}).

\subsection{Nilpotence}

While twisted cohomology using real coefficients appears frequently in the physics literature, twisted homology and cohomology with integral coefficients is rare.  The problem is that, for some choices of $H$, the twisted differentials are not nilpotent, and so the definition (\ref{omrit}) of twisted homology does not make sense, as the image of $\partial_H$ is not a subgroup of the kernel.  

We have assumed that $H$ is a cocycle so $\delta H=0$.  Therefore if $H$ is an odd-degree cochain then the square of the twisted boundary operator is
\bea \label{dquad}
\partial_H^2 x&=&\partial^2 x+\partial (H\cap x)+H\cap(\partial x) + H\cap (H\cap x))\\
&=&0+(\delta H)\cap x-H\cap(\partial x)+H\cap(\partial x)+(H\cup H)\cap x=(H\cup H)\cap x.\nonumber
\eea
On the other hand if the degree of $H$ is even then the action of $H$ on a $p$-chain needs to be multiplied by a factor of $(-1)^p$ to find the same cancellation.  Physically this corresponds to the consistent choice of sign in the worldvolume couplings of $M$-branes to the twist $G_4$.

Eq.~(\ref{dquad}) implies that twisted homology, and similarly twisted cohomology, are only defined when
\beq
H\cup H =0.
\eeq
There are backgrounds in which the cup product of the twist with itself is nonzero.  In type IIB string theory it was conjectured in Ref.~\cite{DMW} that $H\cup H$ contributes to an anomaly.  In Ref.~\cite{aussy2005} the Freed-Witten anomaly was used to demonstrate that this anomaly is canceled by a background D3-brane charge.  This background charge ruins the twisted homology classification.  In M-theory the twist is $G_4$ and the Bianchi identity
\beq
G_4\wedge G_4=d*G_4
\eeq
implies that there is a background M2-brane charge, again ruining the twisted homology classification.

Consider type II string theory, so $H$ is a degree 3 class.  This implies that the cup product is antisymmetric on $H$, and so
\beq
H\cup H = -H\cup H.
\eeq
Adding $H\cup H$ to both sides one finds
\beq
2H\cup H =0
\eeq
and so $H\cup H$ can only be non-zero if it is a degree 6 $\Z_2$-torsion class in the sixth integral cohomology $\H^6(M)$.  If $M$ is an orientable, connected 6-manifold then $\H^6(M)=\Z$ and so the sixth cohomology does not contain any $\Z_2$ torsion.  Therefore, on an orientable 6-manifold, and in particular on a simply-connected 6-manifold
\beq
H\cup H=0\hsp \partial_H^2=0\hsp \delta_H^2=0
\eeq
and so twisted homology and cohomology are well-defined.  Of course, one can define twisted homology on any manifold by simply only allowing values of $H$ such that $H\cup H=0$.  Mysteriously, twisted K-theory is well-defined when $H\cup H=0$ because the differential
\beq
d_3=Sq^3+H\cup
\eeq
that is used in the Atiyah-Hirzebruch construction of twisted K-theory is nilpotent.  However, due to the shifts in brane charges mentioned above, twisted K-theory apparently only classifies D-branes up to a shift in this case.


\section{Classifying Branes} \label{brane}

\subsection{Anomalous and Unstable Branes}
If a D$p$-brane worldvolume contains a boundary or a nontrivial $H$ flux then either its worldvolume Wess-Zumino terms fail to respect RR gauge invariance or the worldvolumes of open strings ending on the brane suffer from a global Freed-Witten anomaly.  In either case the D-brane is inconsistent.  These inconsistencies can be cured by adding another D$p$-brane whose boundary is the boundary of the original D$p$-brane, resulting in a boundaryless D$p$-brane, and by adding a D$(p-2)$-brane whose boundary is dual to the offending $H$ flux \cite{MMS}.  In the second case, the boundary of the D$(p-2)$ is a magnetic monopole in the worldvolume $U(1)$ gauge theory of the D$p$.  In general, there is a quantum correction to the Freed-Witten anomaly equal to the third Stiefel-Whitney class of the normal bundle of the brane, but this vanishes for orientable branes when $M$ is a six-dimensional, orientable, $spin^c$ manifold.

The D$(p-2)$ may be semi-infinite, extending away from the D$p$ to either a spatial or a temporal infinity.  Consider a case in which the D$(p-2)$-brane extends into the past, where it wraps a $(p-2)$-cycle which is nontrivial in untwisted homology.  Furthermore, imagine that the D$p$-brane wraps a compact $(p+1)$-cycle in spacetime.  Then the Freed-Witten anomaly cancellation described above is a dynamical process in which an apparently stable D$(p-2)$-brane decays.  The possibility of such decays in this context was first noted in Ref.~\cite{DMW}, and was described in the current language in Ref.~\cite{MMS}.  They refer to the D$p$-branes, which have finite lifetimes, as (MMS) instantons.  In Ref.~\cite{KPV} the authors claim that each step in the Klebanov-Strassler cascade \cite{KS} is such a process.

Therefore the Freed-Witten anomaly, combined with RR gauge-invariance, has two consequences.  First, some chains cannot be wrapped by any D-brane without insertions.  Second, sometimes a D-brane wrapping an untwisted cycle in ordinary homology can decay.  Therefore ordinary homology does not appear to classify D-branes, as it includes both inconsistent and unstable branes.  

We claim that consistent branes are classified by twisted homology, and that the unstable branes are twisted boundaries, and so correspond to the trivial twisted homology class.  More precisely, if $\sigma$ is a twisted boundary then there exists a chain $\eta$ such that
\beq
\sigma=\partial_H\eta
\eeq
and a brane wrapping $\sigma$ can decay via an MMS instanton which wraps the chain $\eta$.  

However the chain should not be identified with the D-brane worldvolume.  The chain is the sum of the worldvolume not only of the D$p$-brane, but also of any lower-dimensional D-branes dissolved inside, which are magnetic fluxtubes and instantons in the worldvolume gauge theory.  For example a D2-brane wrapping an $S^2$ with a magnetic vortex would be represented by a 2-cycle wrapping the $S^2$ and also a 0-chain, which is a point on the $S^2$.  Given a particular D$p$-brane embedding and a particular abelian or nonabelian worldvolume gauge bundle $F$, these lower brane charges are given at the level of differential forms by the product of the Chern character of $F$ and the square root of the A-roof genus of the tangent space to the D$p$-brane, divided by the square root of the A-roof genus of the normal bundle.  In our current formulation we can only determine the full, integral charge in the abelian case, where it is just the first Chern class of the gauge bundle.  Notice that with this definition a configuration in which a D$(p-2)$-brane is dissolved in a D$p$-brane is in the same twisted homology class as a configuration in which it is not dissolved, also the Myers dielectric effect leaves the twisted homology class invariant as the brane's dimension changes \cite{Luca}.  We will sometimes ignore this subtlety for simplicity and imagine that no branes are dissolved.

To test the claim that branes can consistently wrap any twisted cycle, consider a network of D-branes that wraps a twisted cycle $C$
\beq
C=\sum_i n_i \sigma_i\hsp 0=\partial_H C=\partial C+H\cap C.
\eeq
We want to argue that this network is consistent.  First, imagine that the pullback of the $H$ flux to the worldvolume of one of the D$p$-branes $\sigma_i$ is nontrivial.  Then the worldvolume gauge theory will have a nonvanishing monopole charge $H\cap \sigma_i$.  Therefore $\partial_H\sigma_i\neq 0$ because $\partial\sigma_i$ and $H\cap\sigma_i$ are a $(p-1)$-chain and a $(p-3)$-chain respectively, and so cannot cancel each other.  Therefore the D-brane $\sigma_i$ alone does not wrap a twisted cycle, and correspondingly it is not consistent.  However the sum $C$ of all of the D-branes is $\partial_H$ closed, and so there must be another D-brane or stack of D-branes whose twisted boundary cancels that of $\sigma_i$.  Imagine for simplicity that the twisted boundary is canceled by a single D-brane, with worldvolume $\sigma_j$
\beq
-H\cap\sigma_i=\partial\sigma_j+H\cap\sigma_j.  \label{cancella}
\eeq
The left hand side is a $(p-3)$-chain, so the right hand side must also be a $(p-3)$-chain.  The two terms on the right are of different degrees, and so one must vanish, or equivalently cancel with a contribution from another brane.

First consider the case in which $\partial\sigma_j$ vanishes.  In this case $\sigma_j$ is a $p$-cycle and the brane is another D$p$-brane, with opposite magnetic charge to that of the brane wrapping $\sigma_i$.  Not only is the magnetic charge the opposite, but $-H\cap\sigma_i$ and $H\cap\sigma_j$ are the same chain, and so the monopole and antimonopole are coincident, and therefore there is no source of magnetic charge, only a flux tube.  This flux tube does not violate RR gauge-invariance.  On the other hand if the two branes are separated then the cycles $-H\cap\sigma_i$ and $H\cap\sigma_j$ will no longer be equal, but only homologous, and so $C$ will be $\partial_H$ closed if a D$(p-2)$-brane extends between the monopole-antimonopole pair.  In this case the worldvolume gauge theory of the branes wrapping $\sigma_i$ and $\sigma_j$ will have a $U(2)$ gauge symmetry broken to $U(1)\times U(1)$, and the D$(p-2)$-brane will be a 't Hooft-Polyakov monopole in this theory.  The theory will be consistent, and, thanks to the Wess-Zumino terms of the D$(p-2)$-brane, the sum of all of the Wess-Zumino terms will be RR gauge-invariant.

Now consider the case in which $H\cap\sigma_j$ vanishes in Eq.~(\ref{cancella}).  The D$(p-2)$-brane that we just described is an example of such a situation.  Now
\beq
-H\cap\sigma_i=\partial\sigma_j
\eeq
which means that the brane wrapping $\sigma_j$ is a D$(p-2)$-brane, which finishes on the monopole.  If it is compact it will yield a finite-energy monopole, like the 't Hooft-Polyakov monopole or more generally a monopole of the type introduced by Verlinde in Ref.~\cite{verlinde}.  If it is noncompact it will yield a Dirac monopole.  Depending on the physical question that one is trying to answer, one may wish to accept or reject such configurations, correspondingly one needs to decide whether the embeddings of the simplices are allowed to extend to infinity.  In either case, the D$(p-2)$-brane restores RR gauge-invariance.  

Finally we are left with the case in which some $k$-chain $\sigma_i$ has a nontrivial boundary.  The $\partial_H$ closure of $C$ now implies that this boundary is either canceled by the boundary of another $k$-chain, in which case there is no boundary in the combined chain, or else it is a magnetic monopole on the worldvolume of a $(k+2)$-chain.  In either case the network $C$ wrapping an arbitrary twisted cycle is consistent.

\subsection{Example: The SU(2) WZW Model}
Consider type II string theory on $S^3\times M^7$ where we will not be interested in $M^7$, but will imagine that it contains a time direction along which our branes are extended.  Consider branes not extended along any of the other directions in $M^7$.  Imagine that the $S^3$ supports $k$ units of $H$ flux, corresponding to the level $k-2$ $SU(2)$ WZW model.  The spectrum of D-branes in this WZW model has been computing using conformal field theory, stable branes are even-dimensional and they carry charges in the group $\Z_k$.  Now we will compute the twisted homology of $S^3$ with $k$ units of $H$ flux and compare the result with these expectations.

We can construct $S^3$ using two 3-simplices $\sigma_N$ and $\sigma_S$, which are mapped to the northern and southern hemispheres.  All of their four 2-faces $\sigma^i_2$ are glued together on the $S^2$ equator.  To define the twisted boundary operator we need to determine a particular cocycle which represents the flux $H$.  Different choices of cocycle will yield different twisted boundary operators and so different twisted cycles, but the resulting twisted homologies will be isomorphic.  This is analogous to the fact that the cohomology class of the $H$ flux does not determine the distribution of the $H$ flux, which is also an observable.  We will represent the $H$ flux by a cochain which maps $\sigma_N$ to $k$ times a particular $0$-chain $\sigma_0$ and maps $\sigma_S$ to $0$\beq
\partial_H \sigma_N=k\sigma_0+\sum_i\sigma^i_2\hsp \partial_H \sigma_S=-\sum_i\sigma^i_2.
\eeq
Notice in particular that the sum of $\sigma_S$ and $\sigma_N$ is a 3-cycle but it is not a twisted 3-cycle because
\beq
\partial_H(\sigma_S+\sigma_N)=H\cap(\sigma_S+\sigma_N)=k\sigma_0.
\eeq  

As it is a 3-cocycle, $H$ automatically annihilates all of the $0$, $1$ and $2$-simplices.  This means that on all of the simplices of dimension less than 3, the twisted boundary operator is the ordinary boundary operator, and so the twisted $0$-cycles, $1$-cycles and $2$-cycles are just the untwisted cycles.  In addition the image of $H\cap$ consists entirely of multiples of $k\sigma_0$, and so the twisted $1$-boundaries, $2$-boundaries and $3$-boundaries are just the untwisted $1$-boundaries, $2$-boundaries and $3$-boundaries.  Therefore the degree 1 and 2 chains give the same contribution to the twisted homology as to the ordinary homology, nothing.  Physically this reflects the fact that all of the consistent 1 and 2-branes wrap closed loops and Riemann surfaces which are all contractible in $S^3$, and so the branes can disintegrate.  However we must remember that if a 2-brane has a magnetic vortex, the vortex will correspond to a 0-chain which is not necessarily trivial.

Only the 3-simplices $\sigma_S$ and $\sigma_N$ and the 0-simplex $\sigma_0$ will be relevant, all other 0-simplices are homologous to $\sigma_0$.  Therefore we will only be interested in chains which are linear combinations of these three simplices.  Acting with the twisted boundary operator on such a chain $C$ of the most general form we find
\beq
\partial_H C=\partial_H(n_S\sigma_S+n_N\sigma_N+n_0\sigma_0)=(n_N-n_S)\sum_i\sigma^i_2+kn_N\sigma_0
\eeq
therefore $C$ is a twisted cycle if and only if $n_S=n_N=0$.  The group of relevant twisted cycles is then $\Z$, which is parametrized by $n_0$, and all twisted cycles are of even degree.  Correspondingly the odd degree chain $\sigma_S$ cannot be wrapped because it has a boundary, and the sum of $\sigma_S$ and $\sigma_N$ cannot be wrapped because it has a nontrivial $H$ flux and so a space-filling brane would not respect RR gauge-invariance.  

The twisted boundaries, in dimensions $0$ and $3$, are all multiplies of $kn_N\sigma_0$, which is an index $k$ subgroup $k\Z$ of the group $\Z$ of twisted cycles.  This means that a stack of $k$ D0-branes is unstable.  Consider, for example, $k$ D0-branes at the north pole.  These can blow up into a D2-brane via the dielectric effect, which then carries the D0 charges because it has $k$ magnetic vortices on its worldvolume.  The spherical D2 can sweep out the 3-sphere, but in the process the pullback of the $H$ flux translates into an ever decreasing $B$ field on the D2 worldvolume, which causes the gauge-invariant quantity $B+F$ to decrease from $k$ to zero, when the cycle has been swept.  At this point, one can perform a large gauge transformation
\beq
B\rightarrow B+\Lambda\hsp F\rightarrow F-\Lambda
\eeq
to make $B$ continuous when the brane shrinks away.  This large gauge transformation sets both $B$ and $F$ to zero, and so the vortices disappear.  Thus the D2 can freely shrink into oblivion at the south pole, and the $k$ D0-branes will not reappear.

Now we can take the quotient of the twisted cycles by the twisted boundaries to obtain the twisted homology.  First, notice that there are no twisted cycles, except for those that we have already excluded because they are ordinary boundaries, in odd dimensions.  Therefore the odd twisted homology vanishes
\beq
\HO(S^3)=0
\eeq
in line with the fact that there are no stable odd-dimensional branes in the $SU(2)$ WZW model.  The even-dimensional twisted homology is nontrivial
\beq
\HE(S^3)=\frac{\{n_0\sigma_0\}}{\{kn_N\sigma_0\}}=\frac{\Z}{k\Z}=\Z_k
\eeq
which is consistent with the fact that there are $k$ even-dimensional symmetric boundary states in the WZW model.
If we are willing to consider semi-infinite branes that extend into an $M^7$ direction, then we may add semi-infinite D1-branes that end on a D3-brane wrapping $\sigma_S+\sigma_N$, in other words, wrapping the whole 3-sphere.  This is also a twisted cycle, if we allow vertices of simplices to be mapped to infinity, as $\partial$ of the D1-brane cancels $H\cap$ of the D3-brane.  It is a generalization of the baryon in Ref.~\cite{BBA}.

\subsection{The Spectral Sequence and Multistep Decays}

There is a much easier way to compute the twisted homology of a space.  Twisted homology can be calculated via a series of approximations $E^i_{\rm even}$ and $E^i_{\rm odd}$ that eventually converges on the answer, at least modulo an extension problem.  At each step one defines a differential operator
\beq
d_{2i+1}:E^i_{\rm even}\leftrightarrow E^i_{\rm odd}
\eeq
and the next step in the series is the kernel of this operator divided by its image
\beq
E^{i+1}_{\rm even}=\frac{\Ker({d_{2i+1}:E^i_{\rm even}\rightarrow E^i_{\rm odd}})}{\Im({d_{2i+1}:E^i_{\rm odd}\rightarrow E^i_{\rm even}})}\hsp
E^{i+1}_{\rm odd}=\frac{\Ker({d_{2i+1}:E^i_{\rm odd}\rightarrow E^i_{\rm even}})}{\Im({d_{2i+1}:E^i_{\rm even}\rightarrow E^i_{\rm odd}})}.  
\eeq
This series is called a spectral sequence.  To define it we just need to define $E^0$ and the set differentials.

The first approximation $E^0$ is just the set of even and odd chains.  This is much too large to classify D-branes, intuitively it contains every submanifold with or without boundary.  The first differential is the untwisted boundary operator, and so $E^1$ is just untwisted homology
\beq
d_1=\partial\hsp E^1_{\rm even}=\oplus_i H_{2i}\hsp E^1_{\rm odd}=\oplus_i H_{2i+1}.
\eeq
This is the most commonly used classification scheme for D-branes, although as we have argued it is still too big, because for example it includes cycles that support some $H$ flux so the corresponding branes are inconsistent.

The next differential is the cap product with $H$, which is nilpotent because we are only considering $H$ fluxes such that $H\cup H=0$.  The corresponding approximation $E^2$ is called $H$-homology.  This is a much better classifier of D-branes than $E^1$.  Given any $p$-cycle $C_p$ in the $H$-homology, the cap product $H\cap C_p$ is the trivial element of the ordinary degree $p-3$ homology group, which is a subgroup of $E^1$.  This means that $H\cap C$ is the boundary of a $(p-2)$-chain $C_{p-2}$
\beq \label{defcp2}
H\cap C_p=\partial C_{p-2}. 
\eeq
Therefore a D$p$-brane that wraps $C_p$ is not consistent, but it can be rendered consistent via an insertion of a D$(p-2)$-brane that ends on $H\cap C$.  If we do not allow simplices that extend to infinity, then such configurations are still allowed because the D$(p-2)$-brane may wrap $C_{p-2}$, which does not extend to infinity.  So it may seem like the anomaly is cured with nothing extending to infinity, which would imply that $H$-homology classifies D-branes.

The problem with our network of two branes is that $H\cap C_{p-2}$ is not necessarily zero.  If it is not zero then the D$(p-2)$-brane is also inconsistent, and so requires a D$(p-4)$-brane insertion, which may well need to be semi-infinite, and so the network should not be included in our classification scheme.  The D$(p-4)$ does not need to be semi-infinite if it wraps a cycle $C_{p-4}$ such that
\beq
H\cap C_{p-2}=\partial C_{p-4} \label{toda}
\eeq
but it is not guaranteed that such a cycle exists.  Therefore the $H$-homology approximation $E^2$ is not the full twisted homology, it still contains a D$p$-brane which may require D$(p-4)$ brane insertions, just as branes in $E^1$ required D$(p-2)$-brane insertions and branes in $E^0$ required D$p$-brane insertions.  Each step in the spectral sequence treats brane insertions two degrees lower.

To better approximate twisted homology, so as to eliminate the inconsistent D-branes, we need to eliminate $C_p$ if no $C_{p-4}$ exists satisfying Eq.~(\ref{toda}).  In other words, we need to impose that $H\cap C_{p-2}$ is a boundary.  Recall that $C_{p-2}$ is only defined implicitly, via the relation (\ref{defcp2}).  It is well-defined modulo the addition of a any homology cycle, which means that $H\cap C_{p-2}$ is well-defined modulo the cap product of $H$ with a homology cycle, which is an element of the image of $d_3=H\cap$.  However in arriving at the $H$-homology $E^2$ we have already quotiented by the image of $d_3$, and as an element of this quotient $H\cap C_{p-2}$ is well-defined.  Therefore given an element $C_p$ of $H$-homology we may determine whether or not $H\cap C_{p-2}$ is a boundary, as in (\ref{toda}), and if it is not a boundary we want to eliminate $C_p$.  

We can do this by introducing yet another differential $d_5$, such that
\beq \label{d5}
d_5C_p=H\cap C_{p-2}.
\eeq
The right hand side is determined via Eq.~(\ref{defcp2}).  Equations (\ref{d5}) and (\ref{defcp2}) can be combined by introducing some new notation, called a Toda bracket
\beq
d_5C_p=[H,H,C_p]=H\cap C_{p-2}.
\eeq
The Toda bracket is well-defined as an element of $H$-homology, as
\beq
[H,H,C_p]=H\cap \partial^{-1}(H\cap C_p) \label{todadef}
\eeq
where $\partial$ is invertible on $H\cap C_p$ because $C_p$ is an element of the $H$-homology group $E^2$, so $HC_p$ is a boundary in $E^1$.

\begin{figure}[ht]
\begin{center}
\leavevmode
\epsfxsize 11   cm
\epsffile{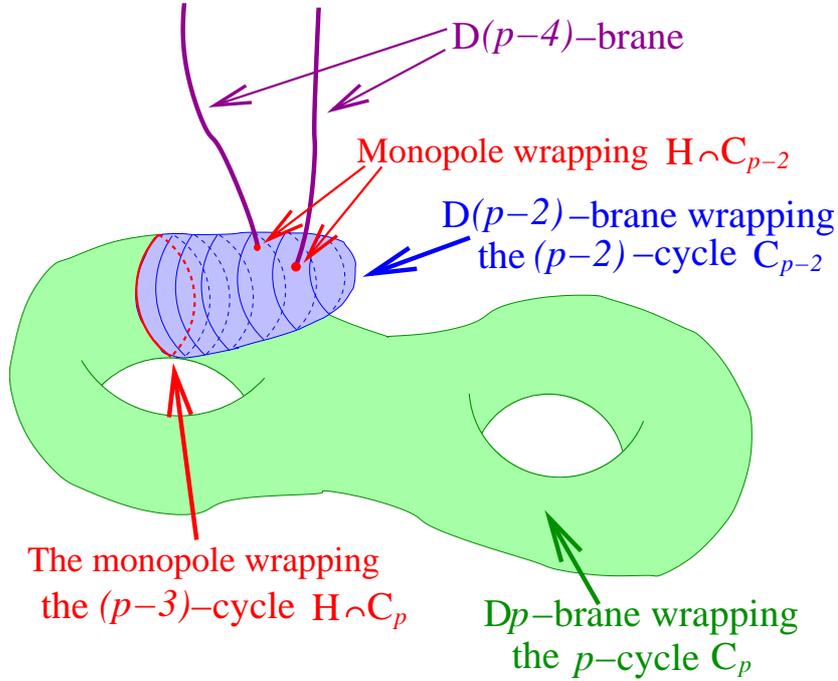}    
\end{center} 
\caption{A D$p$-brane wraps the cycle $C_p$.  The pullback of the $H$ flux to $C_p$ is nontrivial in its worldvolume, although $H\cap C_p$ is a boundary in spacetime.  Therefore the D$p$-brane worldvolume gauge theory has a worldvolume monopole on the cycle $H\cap C_p$.  A D$(p-2)$-brane ends on the monopole, it wraps the chain $C_{p-2}$ which itself supports a nontrivial $H$ flux.  Therefore the D$(p-2)$ also has a monopole, actually two in this picture, on which D$(p-4)$-branes end.}
\label{doppia} 
\end{figure}

We can read the definition (\ref{todadef}) from right to left and interpret each step in terms of the static D-brane network depicted in Fig.~\ref{doppia}.  On the right is our original D$p$-brane $C_p$.  Moving one step to the left, we find its worldvolume monopole $H\cap C_p$, which is the boundary of the D$(p-2)$-brane.  It is a trivial cycle in the spacetime $M$, as $H\cap C_p$ is a boundary, but it is not necessarily trivial as a cycle in the worldvolume $C_p$ of the D$p$-brane.  The worldvolume of the D$(p-2)$-brane is $C_{p-2}=\partial^{-1}(H\cap C_p)$, which itself has a magnetic monopole given by $H\cap\partial^{-1}(H\cap C_p)$.  This in turn is the boundary of a D$(p-4)$ brane, which must extend to infinity if $d_5C_p$ is a nontrivial element of $E^2$.

We can now define $E^3$ to be the kernel of $d_5$ quotiented by the image of $d_5$.  But of course this is not the end, because a cycle in $E^3$ may be wrapped by a D$p$-brane whose anomaly leads to a closed D$(p-2)$-brane whose anomaly leads to a closed D$(p-4)$-brane which itself is anomalous, leading a semi-infinite D$(p-6)$-brane.  To eliminate this possibility one uses a higher Toda bracket
\beq
d_7 C_p=[H,H,H,x]=H\cap\partial^{-1}[H,H,x]
\eeq
which is well-defined as an element of $E^3$.  This process continues until one runs out of dimensions in the compactification manifold.  Notice that if we work over the rationals, and use cohomology instead of homology, then these Toda brackets become the Massey products observed in the spectral sequence for twisted K-theory in Ref.~\cite{AS}, which is the spectral sequence for twisted cohomology in Ref.~\cite{Cavalcanti}.

The spectral sequence is particularly simple in the case of a simply-connected 6-manifold.  For any 6-manifold, all homology classes of degree greater than 6 vanish.  As it is simply-connected, the first and fifth classes also vanish.  Therefore there are no two classes separated by an odd number of degrees greater than three, and so the $H$-homology $E^2$ is already the twisted homology.  Notice that the spectral sequence for twisted K-theory also only has odd differentials, and so it also finishes after $d_3$.  In general $d_3$ for twisted K-theory is not equal to just $H\cap$, it has another term which is a Steenrod square.  However this term will vanish if our 6-manifold is orientable, which it is, because all simply-connected manifolds are orientable.  Therefore we conclude that twisted homology, $H$-homology and twisted K-theory are, as sets, all equal on a six-dimensional simply-connected compactification manifold.

Using this technology we may calculate the twisted homology of $S^3$ with $k$ units of $H$ flux much more quickly than before.  $E^1$ is the ordinary homology
\beq
E^1_{\rm even}=E^1_{\rm odd}=\Z.
\eeq
The operator $d_3$ takes the 3-cycle which generates $E^1_{\rm odd}$ to $k$ times the $0$-cycle which generates $E^1_{\rm even}$.  Therefore, as before, no odd class is in the kernel of $d_3$ and the odd twisted homology is trivial.  Every element of $E^1_{\rm even}$ is a cycle with respect to $d_3$, but multiples of $k$ are in the image.  Quotienting, we find
\bea
E^2_{\rm odd}&=&\frac{\Ker({d_3:E^1_{\rm odd}\rightarrow E^1_{\rm even}})}{\Im({d_3:E^1_{\rm even}\rightarrow E^1_{\rm odd}})}=\frac{0}{0}=0\\
E^2_{\rm even}&=&\frac{\Ker({d_3:E^1_{\rm even}\rightarrow E^1_{\rm odd}})}{\Im({d_3:E^1_{\rm odd}\rightarrow E^1_{\rm even}})}=\frac{E^1_{\rm even}}{kE^1_{\rm even}}=\frac{\Z}{k\Z}=\Z_k\nonumber\\
\eea
as before.  As there are no relative degrees greater than 3 the sequence stops here, and $E^2$ is the twisted homology, as well as the twisted K-theory.

\section{Twisted Homology Classes from Sphere Bundles} \label{sfere}

\subsection{The Bundle}

While the spectral sequence construction of twisted homology is quite abstract, there is a much more concrete construction of twisted homology.  Twisted homology classes in the spacetime $M$ can be realized as ordinary homology classes in a certain bundle
\beq
p:Q\longrightarrow M
\eeq
over the spacetime.  If the twist $H$ is a $k$-cocycle then the fiber is the Eilenberg-MacLane space $\K(\Z,k-1)$, which is defined by the fact that it only has one nonvanishing homotopy group
\beq
\pi_{k-1}(\K(\Z,k-1))=\Z\hsp \pi_{i\neq k-1}(\K(\Z,k-1))=0.
\eeq
Such bundles are entirely characterized by a single characteristic class, which is just $H$.  For example, if $k=2$ then, since $\K(\Z,1)$ is the circle, $Q$ is just a circle bundle and $H$ is its Chern class.

When $k>2$ the Eilenberg-MacLane spaces are infinite-dimensional.  For example, when $k=3$ one finds $\K(\Z,2)$ which is modeled by $\cp^{\infty}$ and also by the projective unitary group on a Hilbert space $PU(\mathcal{H})$, although for a $2k$-dimensional compactification all of the branes lie within a $\cp^{k-1}$ subbundle.  Various relations between this bundle and twisted K-theory are proven in Ref.~\cite{bundlegerbe}.  In the noncommutative field theory truncation of string field theory, there is a $U(\mathcal{H})$ gauge symmetry, whose fundamental fields live in the Hilbert space $\mathcal{H}$.  A single D-brane corresponds to a rank 1 projector at each point, which is a choice of an element of $\cp^\infty$ at each point, in other words a section of the bundle.  Notice that the fact that the characteristic class is $H$, which is topologically trivial when pulled back to a D-branes worldvolume, implies that such a section exists over the worldvolume.  The noncommutative field theory perspective is more powerful than that of the present note as one can consider multiple D-branes, corresponding to higher rank projectors, which yield sections of an infinite-dimensional Grassmannian instead of merely $\K(\Z,2)$.

We will sometimes be interested in a simpler version of $Q$, a $S^{k-1}$-subbundle
\beq
\begin{array}{ccc}S^{k-1} & \longrightarrow & P\\
&&\downarrow\pi\\
&&M\end{array}
\eeq
with Euler class $H$.  In general there may be several such bundles, but any choice will be sufficient for the considerations that follow.  There may also be no such bundle, in which case $P$ does not exist and one needs to work directly with $Q$.  We will prove that when $P$ exists the $H$-homology of $M$ is a quotient of the untwisted homology of $P$, so in particular any twisted cycle is an ordinary cycle in $P$.  We will associate cycles $N$ in $P$ to $p$-branes, where $\pi(N)\subset M$ is their worldvolume and the cycle $N$ wraps the $S^{k-1}$ a number of times equal to the worldvolume $(p+1-k)$-brane charge that comes from magnetic flux tubes.  However, as we will see, not all such cycles in $Q$ may be wrapped.  The cycle $N\subset P$ contains the same data as a generalized submanifold of $M$, in the sense of Refs.~\cite{Gualtieri,Luca1,Luca2}.  We may similarly associate any D-brane to a cycle in $Q$, with lower-dimensional charges corresponding to higher-dimensional cycles wrapped in the fiber $\K(\Z,k-1)$.  These have the advantage that the cycles corresponding to all even dimensional or odd-dimensional D-branes are of the same dimension in $Q$, therefore the twisted even homology and odd homology may each be a single degree homology group in $Q$.  

As the choice of cycle $N\subset Q$ combines the information about the brane's embedding $\pi(N)\subset M$ and its lower $p$-brane charges, it provides a purely geometrical interpretation of features that are difficult to see in $M$.  For example, if a brane wrapping a cycle $\Sigma\subset M$ has a Freed-Witten anomaly, then there are no global sections of $Q$ over $\Sigma$ and the cycle $N$ does not exist.  Branes may decay if and only if the cycle $N$ is the boundary of a chain in $Q$, using the untwisted boundary map.  By comparison, if a brane wraps a nontrivial cycle in $M$ it still may be able to decay via an MMS instanton, whereas if it wraps a trivial cycle in $M$ it may still be stable if it carries lower-dimensional D-brane charges.  Therefore triviality as a cycle in $Q$ is a much better criterion for stability than triviality as a cycle in $M$.  \\

\begin{figure}[ht]
\begin{center}
\leavevmode
\epsfxsize 11   cm
\psfrag{D}{D$3$-brane}
\psfrag{M}{}
\psfrag{Sa}{$S^2_a$}
\psfrag{S}{$S^3$}
\psfrag{d2}{}
\psfrag{d1}{$\pi^{-1}(S^2_a) \longleftrightarrow \partial \tilde P$}
\psfrag{P tilde}{$\tilde P $}
\psfrag{B}{$B_3$}
\psfrag{P}{$P- \pi^{-1}(B_3)$}
\psfrag{A}{D$3$-brane on $S^2 \times \R$}
\psfrag{C}{carrying D$1$ charge}
\includegraphics[width=11cm]{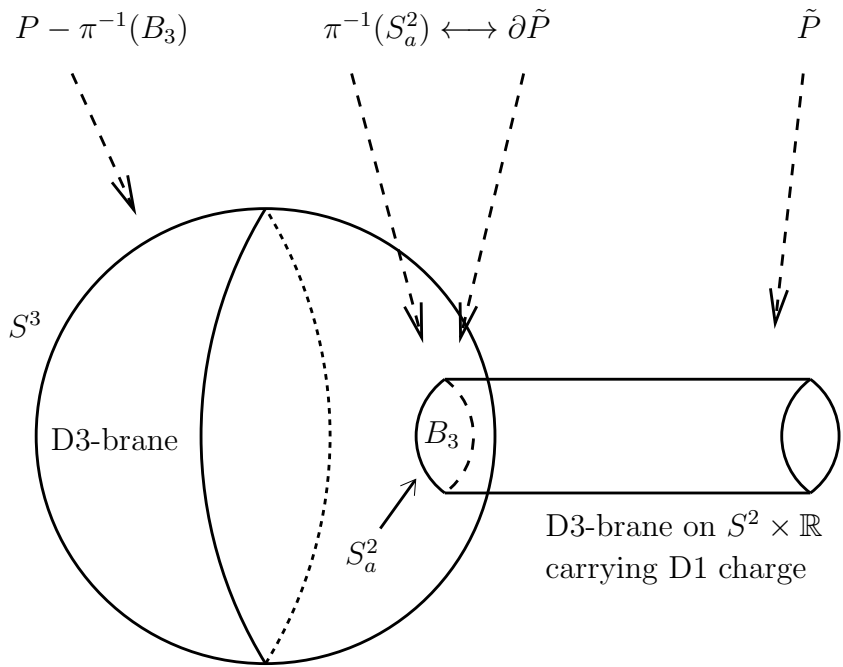}
\caption{Upon excising $B_3$, the total space of the remaining bundle over the D$3$-brane is $P-\pi^{-1}(B_3)$, with boundary $\pi^{-1}(S^2_a)$. The bundle over the `thin' D$3$ on the tube has total space $\tilde P$. A composite bundle is constructed by gluing the respective preimages $\pi^{-1}(S^2_a)$ and $\partial \tilde P$ over $S^2_a$.}
\end{center} 
\end{figure}

We now return to the familiar example of string theory on $M=S^3\times\R^{1,1}$ with $k\neq 0$ units of $H$ flux.  Now $Q$ is a $PU(\mathcal{H})=\cp^\infty$ bundle on $S^3\times\R^{1,1}$ with Dixmier-Douady class equal to $k$ times the top class of $S^3$.  All branes wrap 3-cycles in $Q$.  D3-branes wrapping all of $M=S^3$ are anomalous, corresponding to the fact that there is no global section of $Q$ over $S^3$ and so no available 3-cycle $N$ which projects to the full $S^3$.   There is, however, a global section over $S^3$ minus a small 3-ball $B_3$, where the section over the boundary $\partial B_3 = S^2_a$ of the ball winds around a $S^2=\cp^1\subset\cp^\infty$ subfiber $k$ times.  As there are two spheres in this story, we will name this fiber $S^2_f$.  

Recall that this winding number is just the D1-brane charge. It can be undone as follows. Using the noncompact spatial direction $\R$ one can connect the boundary of the ball with a semi-infinite `tube' $\R^+ \times S^2_a$ that has $S^2_f$ fibered over it, yielding a subbundle with total space
\beq
\tilde P = \R^+ \times S^2 \times S^2_f.
\eeq
At the two sphere $\partial B_3=S^2_a$ where the tube and the anomalous D3 meet, we glue the boundaries of the total spaces of $P - \pi^{-1}(B_3)$ and $\tilde P$ such that the boundary of $\tilde P$ wraps k times the $S^2_f$ fiber over $\partial B_3=S^2_a$. This gluing map is just the transition function of the $PU(\mathcal{H})$ bundle.  The tube corresponds to the $k$ D1-branes that can cancel the anomaly on the D3.  This combined object is a cycle in $[P-\pi^{-1}(B_3)] \bigcup \tilde P$, and it is a valid BIonic D-brane configuration, a baryon in the language of Ref.~\cite{BBA}.

A single D0-brane at a point in $M$ lifts to a 2-cycle $N$ in each timeslice of $Q$ that wraps the $S^2$ subfiber above the point.  Under a slight deformation, $\pi(N)$ is no longer a single point, and so the D0 becomes a D2-brane wrapping a contractible $S^2$ with a unit of D0 charge.  The D0-brane charge is conserved as $N$ generates $\H_2(Q)=\Z_k$.  However a stack of $k$ D0-branes can decay.  This stack lifts to a 2-cycle $N$ that wraps the 2-sphere $k$ times.  If one trivializes $Q$ over the northern and southern hemispheres, then the transition function on the equator takes a cycle that wraps the equator and changes its winding number about the $S^2$ subfiber by $k$ units, in particular it annihilates the winding number of the stack of $k$ D0-branes, reflecting the fact that $k$ is the trivial element of $\H_2(Q)=\Z_k$.  Therefore $k$ D0-branes can decay and correspondingly they wrap a 2-cycle $N\subset Q$ which is a boundary.


\subsection{A Formula for $H$-Cohomology}
In this subsection we will assume that a $S^{k-1}$ sphere bundle exists over $M$ with Euler class $H$.  We will use it to compute the $H$-homology of $M$.  Instead of the $H$-homology of $M$ we will actually calculate the $H$-cohomology, which is isomorphic by Poincar\'e duality on a compact, orientable $M$.  We will use cohomology so that we can use a familiar form of a mathematical tool called the Gysin sequence, without reversing all of the arrows.  

The Gysin sequence is a long exact sequence that relates the cohomology of $P$ to that of $M$ and the Euler class $H$.  It is
\beq
...\stackrel{\pi_*}{\longrightarrow} \H^p(M)\stackrel{H\cup}{\longrightarrow} \H^{p+k}(M)\stackrel{\pi^*}{\longrightarrow} \H^{p+k}(P)\stackrel{\pi_*}{\longrightarrow} \H^{p+1}(M)\stackrel{H\cup}{\longrightarrow} \H^{p+k+1}(M)\stackrel{\pi^*}{\longrightarrow}...
\eeq
where $\pi^*$ is the pullback with respect to the projection map $\pi$ and $\pi_*$ is the pushforward, which at the level of forms is fiberwise integration.  We can make the sequence terminate on the left by dividing the leftmost pair by $\H^p(M)$ and its image $H\cup\H^p(M)=\im(H\cup)$ respectively.  $\H^{p+k}(M)$ has now been quotiented by the image of $H\cup$, which by exactness is the kernel of $\pi^*$, therefore $\pi^*$ is now injective and so the truncated sequence is still exact.  We can also make the sequence terminate on the right by replacing $\H^{p+1}(M)$ with
\beq
\Ker{H\cup}:\H^{p+1}(M)\longrightarrow\H^{p+k+1}(M).
\eeq
This kernel is the entire image of $\pi_*$ so the new truncated sequence is still exact on the right.

Now we have reduced the sequence to
\beq
0\stackrel{H\cup}{\longrightarrow} \frac{\H^{p+k}(M)}{\im(H\cup)}\stackrel{\pi^*}{\longrightarrow}\H^{p+k}(P)\stackrel{\pi_*}{\longrightarrow} \Ker(H\cup)\stackrel{H\cup}{\longrightarrow} 0.
\eeq
Notice that $\pi^*$ is now injective.  Now we repeat the same trick, dividing the two leftmost nonzero entries by the leftmost entry, leaving
\beq
0\stackrel{\pi^*}{\longrightarrow}\frac{\H^{p+k}(P)}{\pi^*\left(\frac{{\rm H}{}^{p+k}(M)}{{\rm{Im}}( H\cup)}\right)}\stackrel{\pi_*}{\longrightarrow} \Ker(H\cup)\stackrel{H\cup}{\longrightarrow} 0
\eeq
which yields an isomorphism $\pi_*$ between the only two nonzero entries.  As the pushforward $\pi_*$ is an isomorphism, we may invert it.  Dividing by $\Im(H\cup)$ we obtain the $H$-cohomology $E^2$ on the right side, and so the isomorphism yields our main result
\beq
E^2_{p+1}(M)=\frac{\H^{p+k}(P)}{\pi^*\left(\frac{{\rm H}{}^{p+k}(M)}{{\rm{Im}}(H\cup)}\right)}/\pi_*^{-1}(\Im(H\cup)). \label{teorema}
\eeq
Notice that the quotient with respect to the $\pi_*^{-1}\Im(H\cup)$ must be done second, as $\pi_*$ is only invertible after the other quotient has been performed.  To write the even or odd twisted homologies when $H$ is a 3-cocycle, we need to sum over all of the even or odd degrees of $E^2_{p+1}$.  Surprisingly in examples, such as $M=S^3$, this formula appears to formally calculate the twisted homology of $M$ although $P$ does not exist.

Intuitively Eq.~(\ref{teorema}) implies that the $H$-cohomology of $M$ is just a quotient of the ordinary cohomology of $P$ by that of $M$, however the quotient with respect to $\H^*(M)$ needs to be performed in two steps.  First one quotients with respect to $\pi^*(\H^{p+k}(M)/\im(H\cup))$ and then by the remaining $\pi_*^{-1}\im(H\cup)$.  As a set this is the same as quotienting by all of $\H^*(M)$, however as a group it is somewhat different.  In fact, $\H^*(M)$ is not necessarily even a subgroup of $\H^*(P)$.  In both quotients one needs to embed the part of $\H^*(M)$ to be quotiented into $\H^*(P)$.  In the first case one uses the pullback $\pi^*$, which preserves the degrees of the classes.  In the second one uses the inverse of the pushforward $\pi_*^{-1}$, which augments the degrees by $k-1$.  Thus different elements of the cohomology of $M$ appear with shifts of different numbers of degrees.  This corresponds to the fact that some branes are lifted from cycles in $M$ to sections of $P$ over $M$, like the D2-brane and D3-brane in the $S^3$ example, whereas others, like the D0 and D1-branes, wrap the sphere when they are lifted.  

\subsection{Twisted Homology of a Lens Space}

To see how this technology works, we will consider a less trivial example.  We will find the twisted homology of the 3-dimensional lens space
\beq
M=\frac{S^3}{\Z_4}
\eeq
whose nontrivial integral cohomology classes are
\beq
\H^0(M)=\H^3(M)=\Z\hsp\H^2(M)=\Z_4.
\eeq
Let $e^i$ be the generator of $\H^i(M)$.  We will consider a twist by the 2-cocycle
\beq
H=2e^2\in\Z_4=\H^2(M).
\eeq
As $S^1=\K(\Z,1)$, the bundles $P$ and $Q$ are the same, they are both $S^1$ bundles over $M$ with Chern class $H$.  In particular, $P$ always exists when $H$ is degree 2 and so Eq.~(\ref{teorema}) can always be used to calculate the twisted homology.

The twisted homology of $M$ is most easily calculated using the spectral sequence, which terminates at the $H$-cohomology because $M$ is only 3-dimensional.  To calculate the $H$ cohomology, we need to know the cup products of $H$ with the various cohomology generators
\beq
H\cup e^0=2e^2\hsp H\cup e^2=H\cup e^3=0.
\eeq
Therefore the kernel of $H\cup$ is $\Z^2\oplus\Z_4$, which consists of all linear combinations of $2e^0$, $e^2$ and $e^3$.  The image consists of all even combinations of $e^2$, but since $4e^2=0$ this leaves only $0$ and $2e^2$.  As the degree of $H$ is two, we do not separate the twisted homology into even and odd.  The twisted homology groups are
\beq \label{spec}
\H^H=\frac{\{2n_0e^0+n_2 e^2+n_3e^3\}}{\{2n_2e^2\}}=\frac{\Z^2\oplus\Z_4}{\Z_2}=\Z^2\oplus\Z_2.
\eeq
The nontrivial torsion class is the charge carried by a 1-brane that wraps the nontrivial cycle in the fundamental group $\pi_1(M)=\Z_4$ an odd number of times.  If it wraps twice, it may decay via a 2-brane instanton which wraps all of $M$ and requires two 1-brane insertions.  The 0-brane and also pairs of 3-branes carry an absolutely conserved charged in this system and so their charges are elements of $\Z$.

The stability and anomalies of these branes are most easily understood in the bundle picture.  Let $\pi:P\rightarrow M$ be an $S^1$ bundle over $M$ with Chern class equal to $H$.  We may calculate the cohomology of the 4-manifold $P$ using the Gysin sequence.  The first part of the sequence
\beq
\H^{-2}(M)=0\stackrel{H\cup}{\longrightarrow} \H^{0}(M)=\Z\stackrel{\pi^*}{\longrightarrow} \H^{0}(P)\stackrel{\pi_*}{\longrightarrow} \H^{-1}(M)=0 \label{zeroiso}
\eeq
demonstrates that $\H^{0}(M)$ and $\H^0(P)$ are isomorphic, so $\H^0(P)$ is the group of integers and $P$ is connected.  Similarly the last part
\beq
\H^{4}(M)=0\stackrel{\pi^*}{\longrightarrow} \H^{4}(P)\stackrel{\pi_*}{\longrightarrow} \H^{3}(M)=\Z\stackrel{H\cup}{\longrightarrow} \H^5(M)=0
\eeq
illustrates that $\H^4(P)$ is isomorphic to $\H^3(M)$, and so again is the group of integers.

The middle of the Gysin sequence is more interesting
\beq \label{mezzo}
0=\H^1(M)\stackrel{\pi^*}{\longrightarrow}\H^1(P)\stackrel{\pi_*}{\longrightarrow} \H^0(M)=\Z\stackrel{H\cup}{\longrightarrow} \H^{2}(M)=\Z_4\stackrel{\pi^*}{\longrightarrow} \H^{2}(P)\stackrel{\pi_*}{\longrightarrow} \H^{1}(M)=0
\eeq
The map $H\cup:\H^0(M)\rightarrow\H^2(M)$ is multiplication by $2e^2$, and so its image consists of the $\Z_2$ subgroup of $\H^2(M)=\Z^4$ generated by $2e^2$.  By exactness, this $\Z_2$ is the kernel of the subsequent map, $\pi^*:\H^2(M)=\Z_4\rightarrow\H^2(P)$ therefore the image of this map is $\Z_4/\Z_2=\Z_2$.  As the next entry is zero, $\pi^*$ is surjective and so $\H^2(P)=\Im(\pi^*)=\Z_2$.  To find $\H^1(P)$, notice that the kernel of $H\cup:\H^0(M)=\Z\rightarrow\H^2(M)=\Z_4$ consists of all even integers $2ne^0$.  Therefore these elements are the image of the previous map $\pi_*:\H^1(P)\rightarrow\H^0(M)$, which is injective.  As the even integers form a group isomorphic to $\Z$, we find that $\H^1(P)$ is isomorphic to $\Z$ by injectivity of $\pi_*$.  

Summarizing, we have found most of the cohomology of $P$
\beq
\H^0(P)=\Z\hsp \H^1(P)=\Z\hsp\H^2(P)=\Z_2\hsp\H^4(P)=\Z.
\eeq
We have not yet found $\H^3(P)$.  In fact, the Gysin sequence does not completely determine $\H^3(P)$ in this case, without knowing the actions of the maps.  It appears in the subsequence
\beq
0=\H^{1}(M)\stackrel{H\cup}{\longrightarrow} \H^{3}(M)=\Z\stackrel{\pi^*}{\longrightarrow} \H^{3}(P)\stackrel{\pi_*}{\longrightarrow} \H^{2}(M)=\Z_4\stackrel{H\cup}{\longrightarrow} \H^{4}(M)=0 \label{maps}
\eeq
which allows $\H^3(P)$ to be either $\Z$, $\Z\oplus\Z_2$ or $\Z\oplus\Z_4$.  To find $\H^3(P)$, we will use Poincar\'e duality.  By Poincar\'e duality, it is isomorphic to $\H_1(P)$, which by the universal coefficient theorem contains the free part of $\H^1(P)$, which is $\Z$, plus the torsion part of $\H^2(P)$, which is $\Z_2$.  Therefore
\beq
\H^3(P)=\Z\oplus\Z_2.
\eeq
This result can be used to reverse engineer the maps in (\ref{maps}).  In particular one finds
\beq \label{whoa}
\pi^*:\H^3(M)=\Z\longrightarrow\H^3(P)=\Z\oplus\Z_2:1\mapsto(2,1)
\eeq
which we will need to plug into (\ref{teorema}) momentarily.

Now we are ready to use Eq.~(\ref{teorema}) to calculate the twisted homology of $M$, which is the $H$-homology as $M$ is 3-dimensional.  As the degree of $H$ is two there is no $\Z_2$ grading to preserve, one sums all of the groups together into a single twisted homology group.  Using Poincar\'e duality the twisted homology group is isomorphic to the twisted cohomology group.  

First we need the cohomology of $M$ divided by the image of $H\cup$.  The image of $H\cup$ is just the $\Z_2$ subgroup of $\H^2(M)=\Z_4$, and so the quotient leaves all of the cohomology of $M$ except that $\H^2$ is reduced from $\Z_4$ to $\Z_4/\Z_2=\Z_2$.  This quotient of $\H(M)$ is embedded in $P$ via the pullback $\pi^*$, and $\H^*(P)$ needs to be quotiented by this embedded subgroup.  According to (\ref{zeroiso}), the pullback is an isomorphism between $\H^0(M)=\Z$ and $\H^0(P)=\Z$, therefore the quotient of $\H^*(P)$ by $\pi^*\H(M)/\Z_2$ eliminates all of $\H^0(P)$.  In (\ref{mezzo}) we saw that the pullback maps the $\Z_2$ quotient of $\H^2(M)$ surjectively onto $\H^2(P)$, therefore the quotient also annihilates $\H^2(P)$.  Finally in (\ref{whoa}) we saw that the pullback maps $\H^3(M)$ onto the subgroup
\beq
\pi^*(\H^3(M)=\Z)=\{(2n,n)\}=\Z\subset\Z\oplus\Z_2=\H^3(P)
\eeq
which according to (\ref{whoa}) leaves the quotient
\beq
\H^3(P)/\pi^*(\H^3(M))=\H^2(M)=\Z_4.
\eeq
The quotient has left $\H^1(P)=\Z$ and $\H^4(P)=\Z$ untouched.  Summarizing, we have found
\beq \label{somm}
\frac{\H^{*}(P)}{\pi^*\left(\frac{{\rm H}{}^{*}(M)}{{\rm{Im}}(H\cup)}\right)}=\Z^2\oplus\Z_4.
\eeq

Finally we need to quotient by $\pi_*^{-1}(\Im(\H\cup))$.  The image of $H\cup$ consists of $0$ and $2e^2$. The preimage of $2e^2$ under the pushforward map $\pi_*:\H^3(P)\rightarrow\H^2(M)$ is the subgroup
\beq
\{2k,n\}\subset\Z\oplus\Z_2=\H^3(P)
\eeq
which, after the quotient by the pullback of the cohomology of $M$, is the subgroup $\Z_2\subset\Z_4$.  Therefore the twisted homology is the quotient of (\ref{somm}) by $\Z_2$ which is
\beq
\H^H(M)=\frac{\Z^2\oplus\Z_4}{\Z_2}=\Z^2\oplus\Z_2
\eeq
in agreement with the much simpler spectral sequence calculation in Eq.~(\ref{spec}).

\section* {Acknowledgements}
We are very greatful to V. Mathai for correspondences, and to L. Martucci for discussions.  This work is supported in part by the Federal Office for Scientific, Technical and Cultural Affairs through the ``Interuniversity Attraction Poles Programme -- Belgian Science Policy" P5/27 and by the European Community's Human Potential Programme under contract
MRTN-CT-2004-005104 ``Constituents, fundamental forces and symmetries of the universe''.


\end{document}